The influence of infant-directed speech on 12-month-olds' intersensory perception of fluent speech

Claudia Kubicek[1], Judit Gervain[2,3], Anne Hillairet de Boisferon[4], Olivier Pascalis[5], Hélène Lœvenbruck[5], and Gudrun Schwarzer[1]

1-Department of Developmental Psychology, Justus-Liebig-University Giessen, Germany

2- Université Paris Descartes, Sorbonne Paris Cité, Paris, France.

3- Laboratoire Psychologie de la Perception, CNRS UMR 8158, Paris, France

4-Department of Psychology, Florida Atlantic University, Boca Raton, United States

5-Laboratoire de Psychologie et Neurocognition, CNRS UMR 5105, Université Pierre Mendès France, Grenoble, France

Correspondence concerning this article should be addressed to Claudia Kubicek, Abteilung für Entwicklungspsychologie, Justus-Liebig-Universität Giessen, Otto-Behaghel-Str. 10 F1, 35394 Giessen, Germany; E-Mail: claudia.kubicek@psychol.uni-giessen.de

# Abstract

The present study examined whether infant-directed (ID) speech facilitates intersensory matching of audio-visual fluent speech in 12-month-old infants. German-learning infants' audio-visual matching ability of German and French fluent speech was assessed by using a variant of the intermodal matching procedure, with auditory and visual speech information presented sequentially. In Experiment 1, the sentences were spoken in an adult-directed (AD) manner. Results showed that 12-month-old infants did not exhibit a matching performance for the native, nor for the non-native language. However, Experiment 2 revealed that when ID speech stimuli were used, infants did perceive the relation between auditory and visual speech attributes, but only in response to their native language. Thus, the findings suggest that ID speech might have an influence on the intersensory perception of fluent speech and shed further light on multisensory perceptual narrowing.

2

# 1. Introduction

Infants are born into a social world and grow up as social beings. In this social world they experience human speech from the beginning. Infants are especially fascinated by speech sounds specifically addressed to them, that is, infant-directed (ID) speech. ID speech is typically characterized by a variety of linguistic and prosodic modifications that are often assumed to facilitate infants' speech processing. Speech processing typically involves the matching of information provided by at least two sensory systems: hearing and vision. The remarkable ability to audio-visually match vowels or syllables (i.e., short speech segments) has been found in infants before 6 months of age (Kuhl & Meltzoff, 1982; Mackain, Studdertkennedy, Spieker, & Stern, 1983; Patterson & Werker, 1999, 2003; Pons, Lewkowicz, Soto-Faraco, & Sebastian-Galles, 2009). Despite the fact that infants experience face-to-face communication mostly in a fluent sequence of utterances, to date, only few studies have focused on infants' intersensory perception of fluent speech, and provided mixed results, in particular in older infants. While one study found 10- to 12-month-olds to perceive the correspondence between native audio-visual fluent speech only (Lewkowicz & Pons, 2013), another study could not find any evidence for this pattern either for native, or for non-native speech in infants at that age (Kubicek et al., 2013). As the two studies differ in the speaking style used, that is, ID speech in Lewkowicz and Pons' (2013), but not in Kubicek et al.'s (2013) work, we assume that ID speech might have facilitated infants' matching performance. However, the speculated contribution of ID speech to 12-month-olds' intersensory speech perception is still unclear and has not yet been addressed. Thus, the motivation of the present study is to investigate whether ID speech does indeed facilitate 12-month-olds' intersensory perception of native and non-native fluent speech.

## 1.1. Characteristics of ID speech

Adults who communicate with infants automatically tend to modify their speech in certain characteristic ways (Fernald et al., 1989) resulting in a distinct speaking style referred to as ID speech. Although cross-linguistic and cross-cultural differences have been observed in the prosodic modifications associated with ID speech (Ingram, 1995), ID speech is generally characterized by an exaggerated prosody (Fernald & Simon, 1984; Uther, Knoll, & Burnham, 2007). In many of the languages that have been studied, ID-speech is characterized by an increase in pitch mean and pitch range relative to adult-directed (AD) speech. Additionally, while ID speech sentences tend to be shorter and often grammatically more simplified or more incomplete than AD sentences (Newport, Gleitman, & Gleitman, 1977), they are slower in tempo, separated by longer pauses, and contain more repetitions of words and phrases (Fernald & Morikawa, 1993; Papousek, Papousek, & Haekel, 1987; Stern, Spieker, Barnett, & Mackain, 1983). From birth on infants prefer to listen to ID speech (Cooper & Aslin, 1990; Fernald, 1985; McRoberts, McDonough, & Lakusta, 2009; Werker, Pegg, & Mcleod, 1994) suggesting that ID speech effectively elicits and holds infants' attention. Likewise, the facial movements made during ID speech differ from AD speech (Chong, Werker, Russell, & Carroll, 2003; Shepard, Spence, & Sasson, 2012) and display characteristics such as exaggerated lip movement (Green, Nip, Wilson, Mefferd, & Yunusova, 2010), exaggerated smiles, raised eyebrows, and wide eyes (Swerts & Krahmer, 2010; Werker & Mcleod, 1989) which are highly salient to infants. These dynamic visual properties that accompany the acoustic ID speech message may further capture infants' attention and may not only facilitate infants' language acquisition (Golinkoff & Alioto, 1995; Graf Estes & Hurley, 2013; Ma, Golinkoff, Houston, & Hirsh-Pasek, 2011; Singh, Nestor, Parikh, & Yull, 2009; Zangl & Mills, 2007) but may also contribute, in particular, to infants' processing of intersensory speech. Although this assumption is speculative, as it is still understudied, a study by Kaplan, Jung, Ryther, and Zarlengo-Strouse (1996) provided some evidence for this hypothesis as they found that 4-month-olds learned an intersensory association between speech segments and faces under an ID speech but not an AD speech condition. Moreover, a recent study by Kim and Johnson (2014) showed that 3- and 5-month-old infants preferred to look at a face talking in an ID speaking style over a face speaking AD speech, highlighting the influence of ID speech on infants' visual attention to talking faces.

**1.2. Infants' intersensory perception of speech**

With respect to the perception of the joint information carried by audio and visual speech streams, infants as young as 2 months (e.g., Patterson & Werker, 2003), and 4.5- to 6-month-olds (Kuhl & Meltzoff, 1982; MacKain, Studdertkennedy, Spieker, & Stern, 1983; Patterson & Werker, 1999) have been found to match audio and visual stimuli corresponding to specific vowels and disyllables in an intermodal matching task (Spelke, 1979). Here, infants are typically presented with two side-by-side video images of a person silently articulating in synchrony, for instance, the vowels /i/ and /a/ while the corresponding sound of one vowel is simultaneously played through a central speaker. If infants exhibit longer looking times to the audio-matching visible speech, cross-modal matching is successfully detected. By using a variant of the intermodal matching task, Pons et al. (2009) demonstrated that 6-month-old infants were able to perceive the intersensory coherence of audio-visual syllables presented sequentially, suggesting that they were capable of performing cross-modal transfer of information. In this kind of paradigm, sequential presentation of auditory and visual stimuli rules out the possibility that infants may simply detect sound-face matching based on audio-visual synchrony, that is, on purely temporal grounds (simply knowing that an open mouth produces louder sounds, for instance), without using more language-based processing mechanisms. By using the same paradigm, a recent study by Lewkowicz and Pons (2013) addressed the question of whether infants are also able to match audio-visual *fluent speech*. The authors tested English-learning infants' ability to orient toward side-by-side silent videos of a Spanish-English bilingual woman talking in a *highly prosodic style* (or ID speech). Spanish was presented on one side and English on the other side before (baseline condition) and after (test condition) auditory-only familiarization with one of the two languages, respectively. The study revealed that 10- to 12-month-old infants showed a novelty preference for the non-native (Spanish) visual speech after they were auditory-only familiarized to their native language (English). The authors assume that infants only recognized the amodal identity of their native language, as a result of perceptual narrowing, that is, a decline in the sensitivity to non-native sensory inputs (for recent reviews, see Lewkowicz, 2014; Maurer & Werker, 2014). Moreover, Lewkowicz and Pons also indicate some audio-visual matching ability in 10- to 12-month-old infants, at least in response to infants' native speech. However, this finding is in contrast with Kubicek et al. (2013) who found by using a nearly identical experimental procedure that 12-month-old German-learning infants failed to show any visual preferences for either German or French silently-speaking faces after auditory input. Methodological issues might have caused those contradictory results, such as, for example, the properties of the languages used, since the contrast between German and French is different in many ways from the contrast between English and Spanish. Importantly, in contrast to Kubicek et al. (2013)[1], Lewkowicz and Pons (2013) used a script that was articulated in a highly prosodic style, that is, in ID speech. Thus, it could be speculated that the 10- to- 12-month-olds' in Lewkowicz and Pons' (2013) study recognized the amodal identity of their native language because the speaking style might have played a role in responsiveness. Due to its highly salient properties, ID speech may have particularly directed infants' attention toward the stimuli. In addition, ID speech may have enhanced the salience of the relevant matching cues (i.e., auditory and facial speech cues) across the modalities, which might have facilitated infants' intersensory response.

### 1.3. The current study

---

[1] In this article the speaking style is described as being child-directed, which refers to the fact that a children's story was used and that the speakers smiled at the beginning and at the end of the video. With respect to facial movements and prosody, these stimuli can be clearly described as being adult-directed.

The present study intended to investigate whether ID speech contributes to 12-month-olds' intersensory matching of audio-visual native and non-native fluent speech.

In Experiment 1, by using Lewkowicz and Pons' (2013) procedure, we presented German-learning infants with visual and auditory speech information presented sequentially in order to control for matching effects due to temporal synchrony. In this task, infants were first exposed to a side-by-side presentation of *silent* videos of one woman talking German on one side and French on the other side (baseline condition). An auditory-only familiarization followed, after which the test started, with the initial silent videos presented again. The German and French stimuli were articulated in AD speaking style. Based on the assumption that infants' looking behavior indicates cross-modal matching, infants were considered to audio-visually match fluent speech if they exhibited longer looking times to the audio-matching visual stimuli during the test trials as compared to the baseline trials.

In Experiment 2, when infants were exposed to ID speech stimuli, we expected the infants to perceive the intersensory correspondence of audio-visual speech by showing a greater proportion of looking time, during the test trials as compared to the baseline trials, toward the visual stimuli corresponding to the audio stimuli they previously heard. According to the perceptual narrowing view, we additionally expected the 12-month-old infants to respond to native auditory speech only. Audio-visual matching with native ID speech would provide evidence for the facilitatory effect of ID speech on intersensory speech perception.

## 2. Experiment 1

### 2.1. Method

#### *2.1.1. Ethics Statement*

The present study was conducted in accordance to the German Psychological Society (DGPs) Research Ethics Guidelines. The Office of Research Ethics at the University of Giessen approved the experimental procedure and the informed consent protocol. Written informed consents were obtained from the infants' parents prior to their participation in the study.

### 2.1.2. Participants

The sample consisted of a total of 49 monolingual German-learning 12-month-old infants ($M_{age}$ = 372 days; $SD$ = 11 days; 22 females). All infants were full-term with no visual or auditory deficits, as reported by parents. The data from 12 additional infants were discarded from the final sample due to extreme fussiness.

### 2.1.3. Stimuli and Apparatus

In Experiments 1 and 2, visual stimuli were silent video clips of two female bilingual German-French speakers. The speakers were recorded against a blue background, looking directly into the camera, while reciting German and French sentences in a similar speaking rate[2]. All videos were resized identically to ensure uniformity. Each of the 30-second video clips showed a full-face image of the speaker and measured 20.6 cm x 18 cm when displayed side-by-side on the monitor, separated by an 11-cm gap. Both videos, French and German, were edited to make sure that the first mouth opening was synchronized. The audio stimuli were the soundtracks extracted from the video recordings (65dB ± 5dB).

Two sets of stimuli were created: adult-directed (Experiment 1) and infant-directed stimuli (Experiment 2).

*Adult-directed (AD) stimuli:* The German and French sentences were adapted from the nursery rhyme "Goldilocks and the three bears" and spoken with a neutral expression in an AD manner, except for the beginning and the end of the story during which the speakers slightly smiled. In both languages, the story consisted of 11 sentences.

### 2.1.4. Procedure

---

[2] Recording took place in Germany, i.e., the bilingual speakers were German-dominant and were rated by French participants (in an informal test) as speaking the French version with a German accent.

Each infant was tested individually in a baby lab. Caregivers gave written informed consent for their infant to participate. Parents were told to keep their eyes closed and to refrain from talking for the duration of the experiment. The infants were seated on the caregiver's lap at a distance of 60 cm in front of a 22-inch monitor. Stimuli were presented using E-Prime 2.0 software (Psychology Software Tools, Sharpsburg, PA).

Following Lewkowicz and Pons (2013), there were six trials (see Figure 1): two silent baseline trials, a first auditory-only familiarization trial followed by a silent test trial during which the two initial silent faces were presented, and a second auditory-only familiarization trial followed by a silent test trial. The left-right position of the videos was counterbalanced across infants, and side of language presentation was switched across baseline and test trials, respectively. In the 3$^{rd}$ and 5$^{th}$ trial (auditory-only familiarization trial) half of the infants were presented with a French voice while watching an attention getter, while the other half heard German. Infants were randomly assigned to one of the two auditory condition groups (German or French).

**Figure 1.** Schematic representation of the procedure used in the current study. Only the speaking style of Experiment 2 (infant-directed speech) and the German auditory condition is shown.

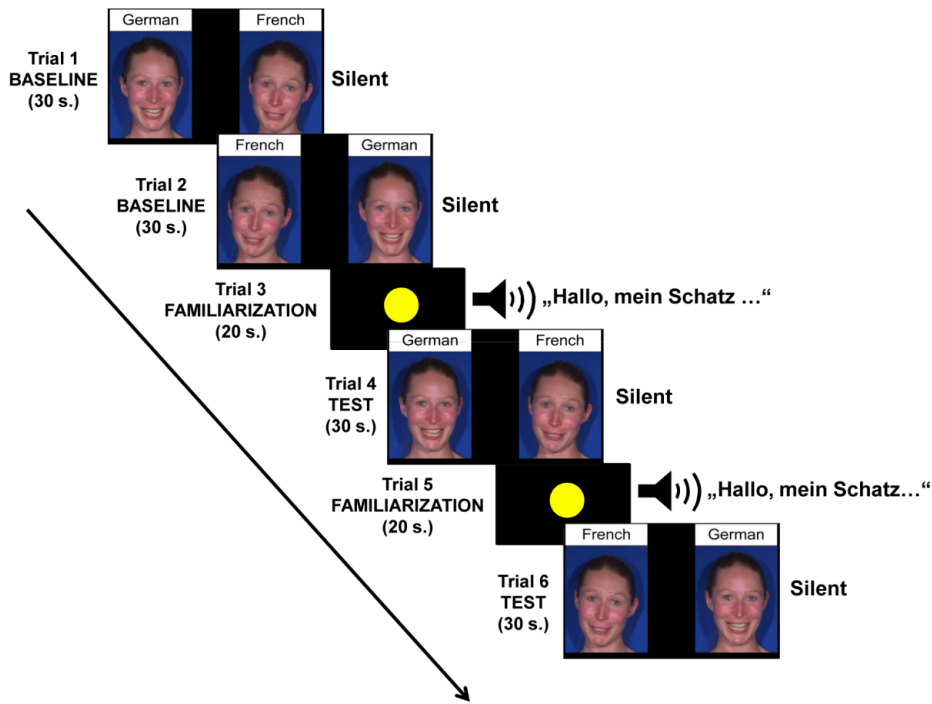

A video camera (specialized for low light conditions) was used to record the infants' looking behavior. The film was then digitized and coded frame by frame by two independent observers who were blind to specific hypotheses. The inter-rater reliability was .92.

**2.2. Results and Discussion**

Following Lewkowicz and Pons (2013), we analyzed looking during the first 20 seconds of baseline and test trials. An analysis of looking times during baseline trials revealed that infants did not show an initial preference for either of the visual stimuli (VS): $M_{GermanVS}$ = 16.0 seconds vs. $M_{FrenchVS}$ = 16.2 seconds, $t(48)$ = -0.22, $ns$.

To determine whether the infants showed intersensory matching, we computed preference scores by dividing the looking time to one face (German talking face or French talking face) by the amount of total looking time (sum of looking times at the German and French talking faces) within baseline and test trials, respectively. We then compared the preference scores for the audio-matching face of baseline to test trials by submitting these scores to a mixed ANOVA with "Trial Type" (baseline, test) as a within-subjects factor and "Auditory Group" (French, German) as a between-subjects factor. The ANOVA revealed no significant differences, indicating that 12-month-old infants did not show any visual preference (Auditory German Group: $M_{Base\_German}$ = 50.1%, SD = 9.3%, $M_{Test\_German}$ = 48.1%, SD = 8.0%; Auditory French Group: $M_{Base\_French}$ = 51.4%, SD = 10.5%, $M_{Test\_French}$ = 50.6%, SD = 11.9%).

Thus, when AD stimuli were used, 12-month-olds did not show an intersensory response to auditory-visual fluent speech when auditory and visual stimuli were presented sequentially. To examine whether ID speech facilitates matching, infants' matching performance was tested in Experiment 2, using the same procedure, with stimuli spoken in an ID style.

## 3. Experiment 2

### 3.1. Method

#### 3.1.1. Participants

The sample consisted of a total of 47 monolingual German-learning 12-month-old infants ($M_{age}$ =369 days; $SD$ = 11 days; 21 females). All infants were full-term with no visual or auditory deficits, as reported by parents. The data from 6 additional infants were excluded due to extreme fussiness.

#### 3.1.2. Stimuli, Procedure and Apparatus

We used the same procedure and apparatus as in Experiment 1 except for the use of ID speech stimuli.

*Infant-directed (ID) stimuli:* For the ID stimuli, we slightly modified the sentences used in Experiment 1 in order to meet most of the characteristics of ID speech. We used 11 sentences, which were *shorter* (half the amount of syllables compared to AD stimuli), had a *simplified grammatical structure*, contained *more repetitions of words and phrases*, and were produced with *a slower tempo* and *longer pauses* (Fernald & Morikawa, 1993; Papousek et al., 1987; Stern et al., 1983). The ID stimuli were constructed to have the same duration as the AD stimuli. The sentences were spoken in a *highly prosodic style*. In both languages, audio ID speech stimuli contained *higher mean fundamental frequency* and *more variation in frequency range* ($M_{German}$ = 268.9 Hz, $SD_{German}$ = 75.0 Hz, $F_0$ $Range_{German}$ = 239.1 Hz; $M_{French}$ = 274.4 Hz, $SD_{French}$ = 61.9 Hz, $F_0$ $Range_{French}$ = 225.5 Hz) than did AD speech stimuli ($M_{German}$ = 220.3 Hz, $SD_{German}$ = 51.6 Hz, $F_0$ $Range_{German}$ = 180.8 Hz; $M_{French}$ = 236.6 Hz, $SD_{French}$ = 48.0 Hz, $F_0$ $Range_{French}$ = 180.5 Hz), mirroring the acoustic differences between ID and AD speech reported in the literature (Fernald, 1989; Fernald & Simon, 1984). The stimuli were also accompanied by facial movements that have been reported to display ID speech characteristics, such as *exaggerated mouth movements, greater teeth visibility, wide eyes, raised eyebrows*, and *smiling eyes and lips* (Chong et al., 2003; Shepard et al., 2012).

The visual German and French ID and AD stimuli were also rated by 22 native German-speaking undergraduate students to assess infant- vs. adult-directedness by using the ID/AD scale developed by Shepard et al. (2012). The scale ranged from 1 (*definitely infant-directed*) to 3 (*neither adult nor infant-directed*) to 5 (*definitely adult-directed*). *Adult-directed* was described as 'the type of speech one uses when talking to an adult, like the way the experimenter is talking to you'. *Infant-directed* was described as 'the type of speech one uses when talking to an infant (a baby between the ages of newborn to one year)'. A score of 0 was used to denote *'I don't know'*. However, none of the participants had chosen the score 0 (*'I don't know'*). In general, the ID speech stimuli were judged to be infant-directed ($M$ = 1.53, $SD$ = 0.25) and the AD speech stimuli were judged as adult-directed ($M$ = 4.17, $SD$ = 0.43), $t(21)$ = 24.81, $p < .001$.

### 3.2. Results

The first analysis revealed that infants did not show a preference for either of the visual speeches during baseline: $M_{GermanVS}$ = 16.9 seconds vs. $M_{FrenchVS}$ = 18.9 seconds, $t(47)$ = -1.68, *ns*.

As in Experiment 1, we computed preferential looking scores, and submitted these scores to a mixed ANOVA with "Trial Type" (baseline, test) as a within-subjects factor and "Auditory Group" (French, German) as a between-subjects factor. The ANOVA revealed a significant Trial Type x Auditory Group interaction, $F(1, 45) = 7.04$, $p < .05$, $\mu^2 = 0.13$, indicating that infants' intersensory matching ability depended on the language they were auditorily familiarized with. Separate two-tailed *t*-tests revealed that infants looked longer at German visual speech after German auditory-only familiarization compared to looking at German visual speech during baseline, $t(24) = -2.79$, $p < .05$, $d = 0.56$. No such difference was found in infants who were auditory familiarized with French, $t(21) = 0.77$, *ns* (see Figure 2).

**Figure 2. Results of Experiment 2.** Mean preference score (%) for German and French visual stimuli during baseline and test conditions. The two sets of bars on the left display data of infants who were auditory-only familiarized with German (native) speech. The four bars on the right side show data of French-familiarized infants. Error bars indicate the standard error of the mean. *Note*. * $p < .05$.

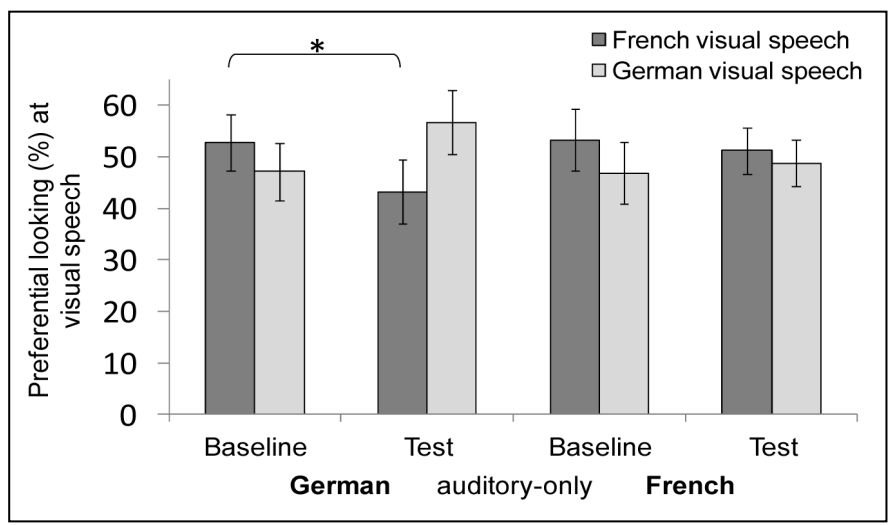

### 3.3. Comparison between Experiment 1 and 2

In order to examine whether speaking style indeed influenced infants' looking behavior, a mixed ANOVA comparing Experiment 1 and 2 with "Trial Type" (baseline, test) as a within-subjects factor, "Speaking Style" (ID, AD), and "Auditory Group" (French, German) as between-subjects factors was conducted. The ANOVA revealed a significant Speaking Style x Trial Type x Auditory Group interaction, $F(1, 92) = 4.56, p < .05, \mu^2 = .05$, indicating that infants' intersensory matching ability depended on the speaking style and the language they were auditorily familiarized with.

Therefore, the findings of Experiment 1 and 2 suggest that ID speech might indeed facilitate 12-month-olds' intersensory matching ability, but only with respect to their native language.

### 4. General Discussion

The principal motivation of the present study was to investigate the influence of ID speech on 12-month-olds' intersensory perception of native and non-native fluent speech. Experiment 1 revealed that infants exposed to AD speech did not show any sensitivity to either German or French visual stimuli in response to the auditory input. In line with our expectation, Experiment 2 revealed that when sentences were spoken in an ID style, infants perceived the intersensory coherence between audible and visible attributes of fluent speech. However, infants' intersensory response was restricted to their *native* language.

The differential efficiency of ID vs. AD stimuli expressed by the fact that infants only showed a matching ability when sentences were spoken in an ID style supports the hypothesis (Fernald, 1984) that ID speech contributes to more efficient information processing in infants. Our results are in line with claims related to the facilitation effect that ID speech is said to have on speech processing (Liu, Kuhl, & Tsao, 2003) and language acquisition (Ma et al., 2011; Singh et al., 2009; Thiessen, Hill, & Saffran, 2005). They are also consistent with prior research on the influence of ID speech on infants' intersensory face-voice associations (Kaplan et al., 1996) and extend these results, as they confirm facilitation of ID speech in infants' intersensory perception of fluent speech (Lewkowicz & Pons, 2013). However, our findings did not exactly meet the results of Lewkowicz and Pons' study (2013). Just as in Pons et al. (2009), infants in the present study looked longer to the audio-matching visual speech. In contrast, the infants in Lewkowicz and Pons' study (2013) showed a novelty preference and looked longer toward non-native visual speech after native auditory language input. Those contradictory results might have been caused by methodological differences, such as the use of different language contrasts, the infants' native language, different age ranges, and stimuli used. More notable than this minor difference, however, is the similarity that both studies observed a preference for one of the visual stimuli in response to *native* auditory-only speech spoken in an ID manner.

The question remains as to why ID style facilitates the processing of auditory-visual fluent speech more than AD style does. One speculation is that the characteristic auditory (or acoustic) and visual (or articulatory) properties of ID speech (Fernald & Kuhl, 1987; Shepard et al., 2012) improve infants' attention to speech cues and are easier to process (Soderstrom, 2007). It has also been shown that ID speech elicits increased neural activity compared to AD speech in 6- and 13-month-olds for familiar words (Zangl & Mills, 2007). Only the older infants (13-month-olds) showed an increased response to ID speech for both familiar and unfamiliar words. These event-related potential data point to developmental changes in the response to ID speech around the first year (Hayashi, Tamekawa, & Kiritani, 2001). Thus, it could be speculated that ID speech enters more into the processing of audio-visual speech than AD speech does, especially in infants toward the end of the first year.

The aforementioned speculation could explain the behavior of the 12-month-olds in Experiment 1, who showed neither visual preference nor auditory-visual matching when exposed to AD speech. Infants toward the end of the first year might be less attentive to visual AD speech cues (Kubicek et al., 2013; Sebastian-Galles, Albareda-Castellot, Weikum, & Werker, 2012; Weikum et al., 2007) and therefore do not notice the differences between the visual cues carried by two different languages. Following this hypothesis, the failure to match audible and visible AD speech could have been caused by the failure to visually discriminate the languages. Less attentiveness to the visual modality of speech is probably caused by an increased understanding of auditory speech in 12-month-olds as, for instance, compared to younger infants entering the babbling stage, who rely more on the redundancies of audio-visual speech (Lewkowicz & Hansen-Tift, 2012). Younger infants at 4 and 6 months of age have been found to be sensitive to visual speech cues (Weikum et al., 2007) and show indeed the ability to match audio-visual fluent speech (Kubicek et al., 2104).

Infants around the first year do not only improve their speech comprehension skills (e.g., for words) by parsing the auditory signal of the speech input (Swingley, 2009) but they are also more and more attentive to facial cues obtaining social information (Brooks & Meltzoff, 2002; Moll & Tomasello, 2004). Moreover, 12-month-olds have been shown to be sensitive to emotional affects in speakers. In this line, it has been shown that vocal-only cues contribute more to social referencing than the visual facial expression of a speaker in 12-month-olds (Baldwin & Moses, 1996; Mumme, Fernald, & Herrera, 1996; Vaish & Striano, 2004). This suggests that infants at this age might be more experienced in processing auditory speech information. Considered in the context that the infants of the present study did not show audio-visual matching due to less attention to the relevant (visual) matching cues, it could thus be speculated that the typical characteristics of ID speech, such as easier syntactic structure, slower speaking rate, and longer pauses as well as exaggerated facial cues (mouth movements, greater teeth visibility, wide eyes, raised eyebrows, and smiling eyes and lips), might have caught infants' attention and finally might have driven 12-month-olds matching performance with regard to their native language.

In Experiment 2, 12-month-olds responded to *native* audio-visual ID speech only. This restriction is in line with the perceptual narrowing/tuning view (Scott, Pascalis, & Nelson, 2007), that is, a decline in the response to non-native input toward the end of the first year found in many perceptual domains such as perception of phonemes (e.g., Werker & Tees, 1984), music perception (Hannon & Trehub, 2005), discrimination of other race (e.g., Kelly et al., 2007) and other species faces (e.g., Pascalis, de Haan, & Nelson, 2002), and audio-visual speech perception (e.g., Kubicek et al., 2014). Furthermore, the results of the present study show continuity with the data of Kubicek et al. (2014) who found that 6-month-old infants were able to perceive the intersensory coherence of *native* audio-visual speech only. Our results thus further support the view that perceptual narrowing is a domain general and pan-sensory process (Lewkowicz & Pons, 2013; Pons et al., 2009).

In conclusion, by conducting two Experiments, the current study provided, for the first time, evidence of the influence ID speech might have on the intersensory perception of fluent speech in 12-month-old infants. Exactly *how* ID speech may have facilitated intersensory perception cannot be determined from the current data and remains a question for future research.

**Funding**

This study was supported through a grant from the German Research Foundation (Deutsche Forschungsgemeinschaft) to GS (SCHW 665/11-1) and ANR--10-FRAL-017 to OP.